\documentclass[10pt,a4paper]{article}
\usepackage[utf8]{inputenc}
\usepackage{amsmath}
\usepackage{amsfonts}
\usepackage{amssymb}
\usepackage{caption}
\usepackage{graphicx}
\begin{document}

\centerline{\LARGE {\bf A mechanism to attract electrons}}
\centerline{Kanchan Meena$^{1}$, P. Singha Deo$^{1}$}
\centerline{$^{1}$S. N. Bose National Center for Basic Sciences, Kolkata, 700106 India}

\noindent{\it Abstract:} In a startling discovery it has been recently found that certain density of states (DOS) can become negative
in mesoscopic systems wherein electrons can travel back in time. 
We give a brief introduction to the hierarchy of density of states in mesoscopic systems as we want to point out some robust phenomenon that can be
experimentally observed with our present day technologies. They can have direct consequences on thermodynamic effects and also can provide
indirect evidence of time travel.
Essentially certain members of the hierarchy of DOS become negative in these regimes and that can attract other electrons.


\bigskip
\bigskip
\bigskip
\bigskip
\bigskip

Mesoscopic systems are so small and subject to such low temperatures that the electronic properties are determined by quantum mechanics while the sample
dimensions compete with the material specific intrinsic scales of the system to produce new physics. While this is the standard definition for a mesoscopic
system, it has become increasingly obvious that to understand these systems we also need to consider the so called leads as an integral part of these systems.
This is essentially because the relevant quantity that determine the relevant electronic properties are the relevant density of states (DOS) and that is connected
to the relevant leads involved in the phenomenon. 
We will explain below 
a hierarchy of DOS that consist of local 
partial density of states, emissivity, injectivity, injectance, emittance, 
partial density of states, and finally the well known local density of 
states and the density of states. Except for the last two there are no 
known analogues for bulk systems. 
We will point out a special phenomenon in this work that will further highlight the role of leads.
Essentially some members of the hierarchy can become negative and that can attract electrons that can be experimentally observed
and gives indirect evidence of time travel.


In the figure 1 we show by the shaded region a typical mesoscopic sample which
has several leads attached to it that are indexed $\alpha$,
$\beta$, $\gamma$, $\delta$, etc. The dimension of the shaded region between the leads is so small that single particle quantum coherence holds
and electron dynamics is governed by Schrodinger equation. 
These lead indices appear in all the formulas to be discussed below showing the
importance of these leads in mesoscopic systems. The $\beta$th lead is
drawn in a special way signifying the tip of a scanning tunneling microscope (STM). The STM tip can have four possible functions. First case is that it does not make
an actual contact and also does not draw or deliver any current but can locally change the electrostatic potential at a point $\textbf{r}$.
Second case is that it does not make a contact but can draw or deliver a current via quantum tunneling. What it means is that the STM tip is weakly
coupled to the states of the sample and exchange of current do not alter the states of the system. Third case is that it makes an actual
contact and becomes like any other lead. Fourth is that it makes a contact and yet does not draw or deliver a current because its chemical
potential is so adjusted. Fourth case is the typical situation of the Landuer-Buttiker three probe conductance set up which is now well established
as a mesoscopic phenomenon greatly studied theoretically as well as experimentally. 
We will mostly analyze the second case with respect to our recent results and show that it is a paradigm to test some of our predictions experimentally.
And then we will also show how these predictions naturally come up also in the well studied case four and not noticed before as the specific design
was not considered.

\begin{figure}[bt]
\centering
\includegraphics[width=.6\textwidth, keepaspectratio]{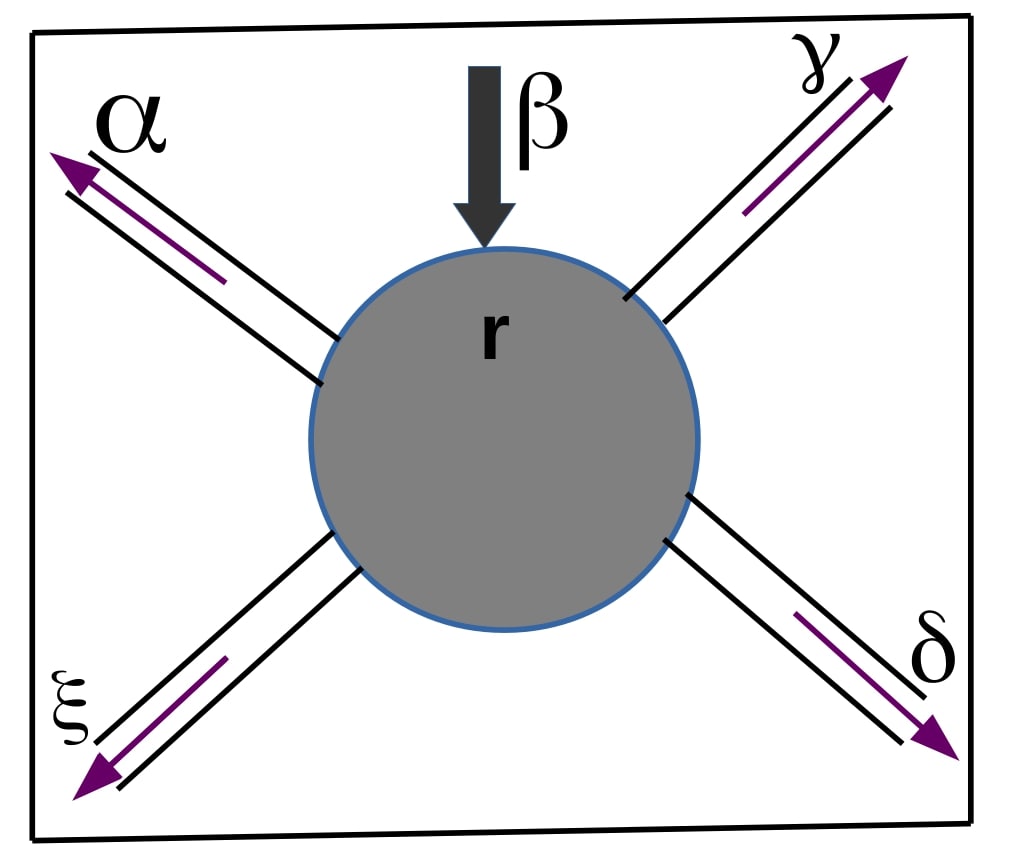}
\captionsetup{labelformat=empty}
\caption{\label{fig3}
Fig. 1 We show a mesoscopic set up where there are many leads $\alpha$, $\beta$, $\gamma$, $\delta$ etc., attached to a sample.
The lead $\beta$ is special in the sense that it is an STM tip that can deliver (or draw) current to (or from) a particular point $\textbf{r}$ in the sample.
All other leads (minimum one other) are fixed and draw (or deliver) current from (or to) the sample.}
\end{figure}


An ensemble of electrons can be incident on the system along the lead $\gamma$ from some classical reservoir (say the terminal of a battery)
and it can be quantum mechanically scattered to any lead wherein one can calculate the scattering matrix element $s_{\alpha\gamma}$ by solving
the Schrodinger equation and applying the relevant boundary conditions. 
The lowest member of the hierarchy is the Larmor precession time $\tau_{lpt}$ for which a detailed derivation can be seen in \cite{deo, but} and given as
\begin{eqnarray}
\tau_{lpt}(E, \alpha, \textbf{r}, \gamma)=-\frac{\hbar}{4\pi i |s_{\alpha\gamma}(E)|^{2}}\left[s_{\alpha\gamma}^*\frac{\delta s_{\alpha\gamma}}
{e\delta U(\textbf{r})}- \frac{\delta s_{\alpha\gamma}^*} {e\delta U(\textbf{r})}s_{\alpha\gamma} \right]
\end{eqnarray}
$\frac{\delta s_{\alpha\gamma}} {e\delta U(\textbf{r})}$ is a functional derivative of 
$s_{\alpha\gamma}$ with respect to the local potential $U(\textbf{r})$ at the point $\textbf{r}$ inside the sample, $E$ is the incident energy 
and $e$ is the electronic charge. 
The ordering of the
arguments on the LHS is important as $s_{\alpha\gamma}$ is a matrix element. Physically it means that an electron going from $\gamma$ to $\alpha$
spends precisely this amount of time at the point $\textbf{r}$. Thus all three indices $\gamma$, $\textbf{r}$ and $\alpha$ are spatial indices and their
ordering is important in all the subsequent formulas we discuss. Since time spent in a propagation is related to states accessed in the process,
both being related to the imaginary part of the Greens function one gets a
local partial density of states (LPDOS) $\rho_{lpd}$ defined as $\frac{|s_{\alpha\gamma}(E)|^{2}}{\hbar} \tau_{lpt}$.
\begin{eqnarray}
\rho_{lpd}(E, \alpha, \textbf{r}, \gamma)=-\frac{1}{4\pi i}\left[s_{\alpha\gamma}^*\frac{\delta s_{\alpha\gamma}}
{e\delta U(\textbf{r})}-\frac{\delta s_{\alpha\gamma}^*} {e\delta U(\textbf{r})}s_{\alpha\gamma} \right]
\end{eqnarray}
The electrons that are involved in going from $\gamma$ to $\alpha$ are $|s_{\alpha\gamma}(E)|^{2}$ in number and these being indistinguishable the factor
$|s_{\alpha\gamma}(E)|^{2}$ in going from Eq. 1 to Eq. 2 is just an averaging over individual electrons.
At zero temperature fermions occupy one state each and for non-interacting Fermions doubling the input flux in $\gamma$ will double the output flux
in $\alpha$ as long as we are not in the completely filled band.
We cannot get linear superposition of states in the input channels that can be argued \cite{deo}.
Numerical simulations \cite{muk}
suggest that Eq. 2 is also
valid in presence of electron-electron interaction in the sample.
Now we can make an averaging over any one or any two or any three of the coordinates $\alpha$, $\textbf{r}$ and $\gamma$ in Eq. 2 and accordingly get
higher members of DOS in the hierarchy. Summing over $\gamma$ means averaging over all incoming channels. Summing over $\alpha$ means averaging
over all outgoing channels. Accordingly different members can be physically interpreted.
Thereby partial DOS is defined as
\begin{eqnarray}
\rho_{pd}(E, \alpha, \gamma)=-\frac{1}{4\pi i}\int_{\Omega}d^{3}\textbf{r}\left[s_{\alpha\gamma}^*\frac{\delta s_{\alpha\gamma}}{e\delta U(\textbf{r})}-
\frac{\delta s_{\alpha\gamma}^*}{e\delta U(\textbf{r})}s_{\alpha\gamma} \right]
\end{eqnarray}
Here $\Omega$ stands for the spatial region of the sample that is the shaded area in figure 1.
Injectivity can be defined as
\begin{eqnarray}
\rho_{i}(E, \textbf{r}, \gamma)=-\frac{1}{4\pi i}\sum_{\alpha}\left[s_{\alpha\gamma}^*\frac{\delta s_{\alpha\gamma}}{e\delta U(\textbf{r})}-
\frac{\delta s_{\alpha\gamma}^*} {e\delta U(\textbf{r})}s_{\alpha\gamma} \right]
\end{eqnarray}
Emissivity can be defined as
\begin{eqnarray}
\rho_{e}(E, \alpha, \textbf{r})=-\frac{1}{4\pi i}\sum_{\gamma}\left[s_{\alpha\gamma}^*\frac{\delta s_{\alpha\gamma}}{e\delta U(\textbf{r})}-
\frac{\delta s_{\alpha\gamma}^*} {e\delta U(\textbf{r})}s_{\alpha\gamma} \right]
\end{eqnarray}
Injectance can be defined as
\begin{eqnarray}
\rho(E, \gamma)=-\frac{1}{4\pi i}\int_{\Omega}d^{3}\textbf{r}\sum_{\alpha}
\left[s_{\alpha\gamma}^*\frac{\delta s_{\alpha\gamma}}{e\delta U(\textbf{r})}-\frac{\delta s_{\alpha\gamma}^*}{e\delta U(\textbf{r})}
s_{\alpha\gamma} \right] 
\end{eqnarray}
Local density of states (LDOS) can be defined as 
\begin{eqnarray}
\rho_{ld}(E, \textbf{r})=-\frac{1}{4\pi i}\sum_{\alpha\gamma}\left[s_{\alpha\gamma}^*\frac{\delta s_{\alpha\gamma}}{e\delta U(\textbf{r})}-
\frac{\delta s_{\alpha\gamma}^*}{e\delta U(\textbf{r})}s_{\alpha\gamma} \right]
\end{eqnarray}
Finally we get DOS where all that can be averaged is summed to give
\begin{eqnarray}
\rho_{d}(E)=-\frac{1}{4\pi i}\sum_{\alpha\gamma}\int_{sample} d^3\textbf{r} \left[s_{\alpha\gamma}^*\frac{\delta s_{\alpha\gamma}}{e\delta U(\textbf{r})}-
\frac{\delta s_{\alpha\gamma}^*}{e\delta U(\textbf{r})}s_{\alpha\gamma} \right]
\end{eqnarray}
\begin{eqnarray}
\rho_{d}(E)=-\frac{1}{2\pi}\sum_{\alpha\gamma}\int_{sample} d^3\textbf{r} \left[|s_{\alpha\gamma}|^{2}\frac{\delta \theta_{s_{\alpha\gamma}}}{e\delta U(\textbf{r})}\right]
\end{eqnarray}
Here $s_{\alpha \gamma}=|s_{\alpha \gamma}|exp(i \theta_{s_{\alpha\gamma}})$.
This is the mesoscopic version of Friedel sum rule that relates scattering phase shift to DOS and does not depend on the lead indices or coordinate as they have
been summed. But there are several lower members that explicitly depend on the lead indices and let us discuss one of them, say $\rho_i(E, \textbf{r}, \gamma)$,
that is injectivity and others can be similarly interpreted. It depends on the input lead index $\gamma$ and physically means the following. 
The quantity applies to only those electrons that are incident along lead $\gamma$. Individual members of this ensemble of electrons may or may not pass through
a remote point $\textbf{r}$ and in fact there is no equation of motion that tells us whether it will. Note that Schrodinger equation
works only for an ensemble and gives us only a probability for it. 
At zero temperature below Fermi energy when we do not distinguish between counting electrons (that constitute a current) and counting states,
$\rho_i(E, \textbf{r}, \gamma)$ give the fraction of those electrons that come from $\gamma$ and 
pass through $\textbf{r}$. Quantum mechanics with its probabilistic interpretations is incapable of saying where these
electrons are going after they enter the sample and only a probabilistic answer can be given.
Yet we can calculate members of this hierarchy and
in case of some further lower members one has to specify to which lead the electron finally goes. This may give the
impression that some of these members are over specified
purely theoretical entities as there is no equation of motion
for answering where does an electron coming from $\gamma$ go after some time.
But we will see that such $\textbf{r}$ dependent DOS matter in experiments.

Injectance is the first member that can be completely specified as it require us to only specify the incoming channel, the rest being summed.
Injectance is total injected current at zero temperature when only lead $\gamma$ bring electrons into the system while all other leads carry electrons
away from the system and $\textbf{r}$ is also integrated out. Injected current is of the form
$nev$ or differential current is $\frac{dn}{dE}evdE$. Electronic charge $e$ can be set to unity and if properly normalized wave-functions are 
taken then we can also drop the $v$ factor \cite{deo} making injected current to be $\frac{dn}{dE}$ at an energy $E$. Now that can be determined from
internal wavefunction $ \psi(\textbf{r}, \gamma)$
when the scattering problem is also set up such that electrons are incident only along lead $\gamma$ and all other leads carry
electrons away from the system so that we do not have to bother where goes the electrons that pass through $\textbf{r}$. That gives
\begin{eqnarray}
\rho(E, \gamma)=\int_{sample} d^3\textbf{r} \sum_{k_{\gamma}}\vert \psi(\textbf{r}, \gamma)\vert ^{2}\delta(E-E_{\gamma,k_{\gamma}})=\\
\int_{\Omega}d^{3}\textbf{r}\sum_{\alpha}-
\frac{1}{2\pi}\left[|s_{\alpha\gamma}|^2\frac{\delta \theta_{s_{\alpha\gamma}}}{e\delta U(\textbf{r})} \right] 
\end{eqnarray}
Eq. 10 is the standard definition of DOS when only lead $\gamma$ bring electrons into the system and the momentum states of these electrons in lead $\gamma$
are determined by the wave vector $k_\gamma$. Lower members of the hierarchy cannot be defined in terms of internal wavefunction
$\vert \psi(\textbf{r}, \gamma)\vert ^{2}$ 
as one can never write down an internal wavefunction that depend on two lead indices and $\textbf{r}$. 
But they can still be defined in terms of the scattering matrix elements or asymptotic wavefunctions far away from $\textbf{r}$.
By appealing to physical process like spin precession and Larmor frequency \cite{deo,but} we can address issues like a particle
going from $\gamma$ to $\alpha$ how much time it spends at the point $\textbf{r}$ and how many (a count or a measure) partial states it occupied at the point
$\textbf{r}$.
We thought wavefunction is the most fundamental entity in quantum mechanics that is determined once we know the internal potential $U(\textbf{r})$
and hence the Hamiltonian. 
We always thought that a state is an entity in Hilbert space. Local density of states can only be defined through ensemble averaging
wherein equal apriory probability implies that all states in Hilbert space are equally accessible by the electrons and time averages give phase space
averages. Averaging over all possible variations in $U(\textbf{r})$ help taking the problem from Hilbert space to
phase space. But say for a benzene molecule attached to leads
if we change the internal potential $U(\textbf{r})$, then it is no longer a benzene molecule. An electron coming from lead $\gamma$ and going to lead
$\alpha$ will not access all states of the benzene molecule but some partial states for which Eq. 3 give partial density of states that cannot be defined
in terms of the internal wavefunction. The integrand in Eq. 10 cannot be broken down into an $\alpha$ dependent quantity.
If we remove the integration over $\textbf{r}$ in Eq. 10 then the integrand does not give any lower member of hierarchy as fixing an $\textbf{r}$ means infinite
uncertainty in momentum and it is not enough to be limited to the momentum state at a particular energy $E$. Likewise, removing the sum over $k_\gamma$ would
mean looking at a particular momentum state and that would mean an infinite uncertainty in coordinate of the electron and so integrating the coordinate over
the sample region does not give anything. Besides a delta function cannot be written unless there is a sum or an integration over its variable.


\begin{figure}[bt]
\centering
\includegraphics[width=.7\textwidth, keepaspectratio]{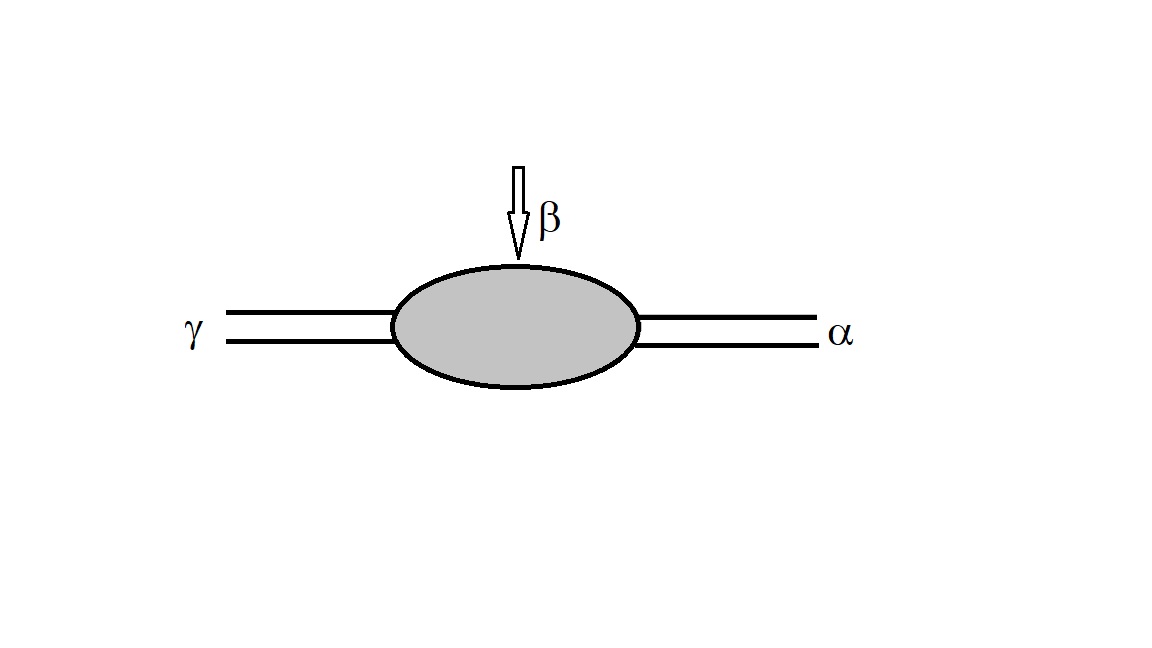}
\captionsetup{labelformat=empty}
\caption{\label{fig3}
Fig. 2 Here we show a simpler version of Fig. 1 as now there are only two fixed leads indexed $\gamma$ and $\alpha$ apart from the STM tip $\beta$. 
This is a cartoon of an experimental set up that help us address some over specified quantities in mesoscopic physics like injectivity and emissivity.
We know in quantum mechanics one cannot experimentally measure coordinate dependent DOS but this set up does allow such measurements indirectly.}
\end{figure}

Now we will show how some of the lower members manifest in experiments.
Consider the situation shown in Fig. 2. We know that quantum states on an infinite 1D line is given by Schrodinger equation with DOS being
$\frac{2}{hv}$ independent of whether these states are occupied by bosons or fermions and independent of temperature. Similarly we say that
the STM tip has a DOS given by $\nu_{\beta}$ and the point $\textbf{r}$ has an over specified DOS.
At zero temperature below the fermi energy 
the transmission probability $T$ of quantum mechanics is $\frac{h}{2e}$ of electronic current
at that particular energy $E$ and that is how the following formulas are to be interpreted.
Let us consider the situation when the tip of $\beta$ is not making a physical
contact with the sample but close enough to deliver (or draw) a current to 
(or from) the sample by tunneling. 
\begin{equation}
T_{\beta \alpha}^e=4\pi^2 \nu_\beta |t|^2 
\rho_{i}(E, \textbf{r}, \alpha)
\end{equation}
\begin{equation}
T_{\alpha \beta}^i=4\pi^2 \nu_\beta |t|^2 
\rho_{e}(E, \alpha, \textbf{r})
\end{equation}
\begin{equation}
T_{\beta \gamma}^e=4\pi^2 \nu_\beta |t|^2 
\rho_{i}(E, \textbf{r}, \gamma)
\end{equation}
\begin{equation}
T_{\gamma \beta}^i=4\pi^2 \nu_\beta |t|^2 
\rho_{e}(E, \gamma, \textbf{r})
\end{equation}
For example, transmission probability $T_{\beta \alpha}^e$ is contribution of lead $\alpha$ to emission current taking place through lead $\beta$ and it is 
proportional to the injectivity
of lead $\alpha$ to the remote point $\textbf{r}$. Others can be similarly interpreted. 
Here $\nu_\beta$ is the density of states in the lead $\beta$ that couple to the states at the point $\textbf{r}$ through the coupling parameter $t$.
Details of this can be found in reference \cite{but}.
Eqs. 12 and 14 correspond to current drawn by the lead $\beta$ while Eqs. 13 and 15 corresponds to that delivered by lead $\beta$.

For the same set up in Fig. 2 with the lead $\beta$ 
not making an actual contact but allowing tunneling to or from the sample, 
we want to address
the current flowing from $\gamma$ to $\alpha$. Series of works by Buttiker \cite{but} give us
\begin{equation}
|S_{\alpha \gamma}^/|^2= |S_{\alpha \gamma}|^2 - 4\pi^2 \mid{t}\mid^2 \nu_\beta \rho_{lpd}(E, \alpha, \textbf{r}, \gamma)
\end{equation}
where $S_{\alpha \gamma}^/$ is the scattering matrix element for scattering from $\gamma$ to $\alpha$ when the STM tip is drawing a current given by the
second term on RHS. When the STM tip is removed by $t\rightarrow 0$ then this scattering amplitude will be $S_{\alpha \gamma}$. Now in quantum mechanics
an electron coming from $\gamma$ can go to the STM tip or to the lead $\alpha$ or can get reflected back
rather randomly and there is no equation of motion for such an electron.
Schrodinger equation is an equation for an ensemble of electrons and gives a probabilistic answer of
$T_{\alpha \gamma}=|S_{\alpha \gamma}|^2$
completely ignoring how the individual electrons are behaving. Now it is easy to translate this problem to statistical mechanics at zero temperature
where the chemical potential of the STM tip as well as that of lead $\alpha$ is set to zero (they are earthed) and there is no return for the electrons that
go there. Chemical potential of $\gamma$ being non-zero will send in an ensemble of electrons. Now in this statistical mechanical problem there will be
an observable current given by $\frac{2e}{h}|S_{\alpha \gamma}^/|^2$.
This measurable quantity therefore directly depends on the local partial density of states.
Transmission probability multiplied by a factor $\frac{2e^2}{h}$ gives the measured conductance.
So $|S_{\alpha \gamma}^/|^2$ and $|S_{\alpha \gamma}|^2$ are both measurable and so in relative proportions $\rho_{lpd}(E, \alpha, \textbf{r}, \gamma)$
is also measurable.
Intuitively, one would think that $\rho_{lpd}$ is positive definite and so the conductance in
presence of the lead $\beta$ is always less than that in the coherent situation.
Recent works show that $\rho_{lpd}$ can be designed to be negative \cite{deo} by creating Fano resonances
and using this set up we can confirm its negativity.

Now in statistical mechanics we can change many parameters and study this observable current. We focus on the situation shown in Fig.3.
Here the reservoirs are explicitly shown as electron reservoirs with definite chemical potentials $\mu_\gamma$, $\mu_\alpha$ and $\mu_\beta$.
\begin{figure}[bt]
\centering
\includegraphics[width=.7\textwidth, keepaspectratio]{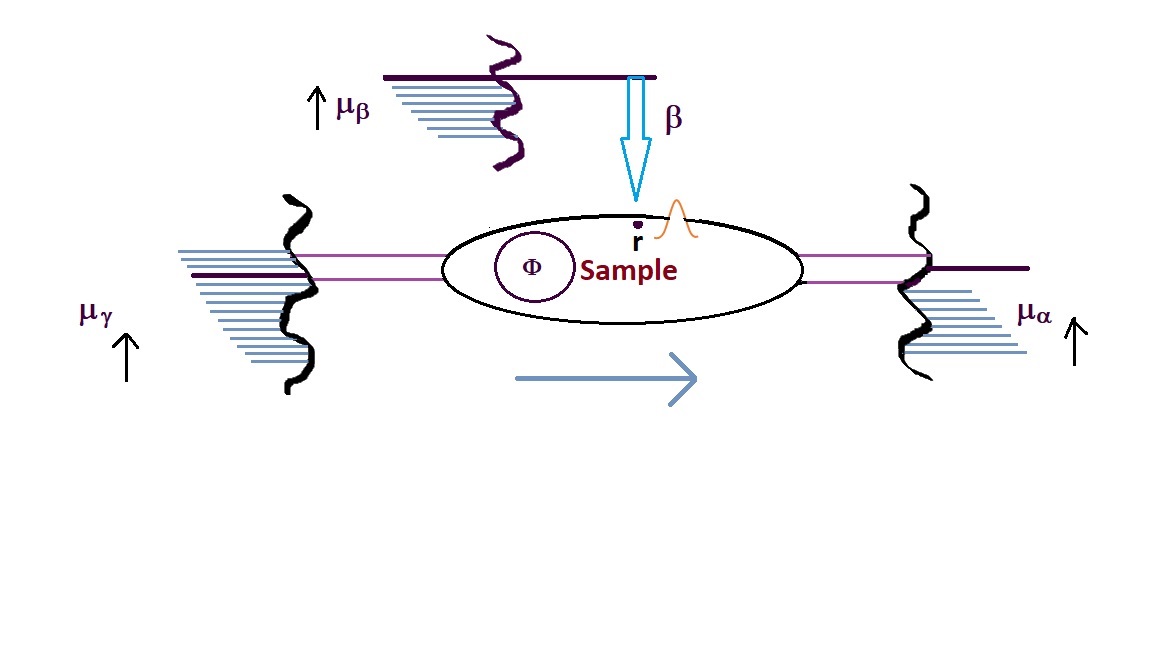}
\captionsetup{labelformat=empty}
\caption{\label{fig3}
Fig. 3 The role of lead $\beta$ (an STM tip) is to inject or remove
electrons from the system as well as causing some decoherence.
The STM tip makes an actual contact with the sample at point $\textbf{r}$ and $\mu_\beta$ is so adjusted that there is no net current
flowing through the STM tip but it can still cause decoherence.
At $0^o$K, a current flows from lead $\gamma$ to lead $\alpha$ in the energy interval $(\mu_\gamma - \mu_\alpha)$.
One can put an Aharonov-Bohm flux $\Phi$ through the sample which is very useful to separate coherent and incoherent effects.} 
\end{figure}
In a situation wherein the probe $\beta$ makes an actual contact with the sample we get
a three probe set up and also the probe $\beta$ is made like a voltage probe in the sense that
its chemical potential is so adjusted that it does not draw any net current from or into the system. This leads to the celebrated Landauer-Buttiker three
probe conductance given by \cite{dat}
\begin{equation}
G=-G_{\alpha \gamma} - \frac{G_{\alpha \beta} G_{\beta \gamma}}{G_{\beta \alpha} + G_{\beta \gamma}}
\end{equation}
Here
\begin{equation}
G_{\alpha \beta}= \frac{2e^2}{h}|S_{\alpha \beta}|^2 \;\;\; for \;\;\; \alpha\ne\beta \;\;\; etc.
\end{equation}
This formula can be rewritten in terms of the hierarchy of the density of states in the following way \cite{but}.
\begin{equation}
G=\frac{2e^2}{h}\Large(|S_{\alpha \gamma}|^2 -4 \pi^2 |t|^2 \rho_{lpd}(\alpha, \textbf{r}, \gamma) + 4 \pi^2 |t|^2 \frac{\rho_e(\alpha,\textbf{r}) 
\rho_i(\textbf{r},\gamma)}{\rho_{ld}(\textbf{r})}\Large)
\end{equation}

Note that in the above formula if the lead $\beta$ is completely removed then $|t|^2=0$ and
we will be left with only the first of the three terms. This is the standard
two probe Landauer conductance formula.
So the three terminal formula of Eqn. (17)
is now restated in the form of Eqn. (19). The second term comes with a negative sign and unless $\rho_{lpd}$ is designed to be negative,
accounts for the loss of coherent electrons
due to the lead $\beta$.
These electrons that loose coherence are not escaping to lead $\beta$ (as $\beta$ is not drawing any net current and $\nu_\beta$ do not affect
this term). This loss affects only those partial electrons
that are going from $\gamma$ to $\alpha$ coherently and hence this term depend on Aharonov-Bohm flux, reducing the overall flux dependence of $G$. 
It is proportional to the local partial density of states at the point $\textbf{r}$ means again it is related to only those
partial electrons going from $\gamma$ to $\alpha$ at the point $\textbf{r}$. 

The lost electrons are momentarily incoherent particles at the point $\textbf{r}$ and eventually redistribute to $\gamma$ and $\alpha$.
The question arises what will be the ratio of this redistribution 
and this will again be determined by the members of the hierarchy. The contribution to $G$ is the third term
separately written below and it is in fact the incoherent contribution in the conductance from $\gamma$ to $\alpha$. 
\begin{equation}
4 \pi^2 |t|^2 \frac{\rho_e(\alpha,\textbf{r}) \rho_i(\textbf{r},\gamma)}{\rho_{ld}(\textbf{r})}
\end{equation}
Note that this term consist of the product of two independent probabilities associated with two separate processes. One involving an injectivity from
$\gamma$ to $\textbf{r}$ and the other involving emissivity from $\textbf{r}$ to $\alpha$.

To understand the denominator in Eq. 20 let us consider the following.
Total number of incoherent electrons at the point $\textbf{r}$ must be
\begin{equation}
4 \pi^2 |t|^2 \frac{\rho_e(\alpha,\textbf{r}) \rho_i(\textbf{r},\gamma)}{\rho_{ld}(\textbf{r})}+
4 \pi^2 |t|^2 \frac{\rho_e(\gamma,\textbf{r}) \rho_i(\textbf{r},\gamma)}{\rho_{ld}(\textbf{r})}
\end{equation}
\begin{equation}
=4 \pi^2 |t|^2 \frac{(\rho_e(\alpha,\textbf{r})+\rho_e(\gamma,\textbf{r})) \rho_i(\textbf{r},\gamma)}{\rho_{ld}(\textbf{r})}
\end{equation}
The first term in Eq. 21 is just the term in Eq. 20 and gives the fraction of incoherent electrons at 
$\textbf{r}$ that goes to $\alpha$ and the second term
is that which goes to $\gamma$. Which means Eq. 22 give the total incoherent electrons at the point $\textbf{r}$. Given the fact that
$\rho_e(\alpha,\textbf{r}) + \rho_e(\gamma, \textbf{r})=\rho_{ld}(\textbf{r})$, Eq. 22 is simply proportional to
$\rho_i(\textbf{r},\gamma)$. This is the quantity that has to be balanced against the chemical potential of the lead $\beta$ at all flux so that lead
$\beta$ does not draw or deliver any net current. This is a situation wherein we are at $0^o$K and in the regime of incident energy $E$ being 
such that $\mu_\gamma > E > \mu_\alpha$. In this regime there is no injectivity from lead $\alpha$.


Strangely enough
if $\rho_{lpd}$ is made negative then the second term in Eq. 19 becomes positive implying the system draws in coherent electrons to the point
$\textbf{r}$ instead of loosing them.
This can be also interpreted as loosing coherent electrons in reverse time.
The same signature can also be seen from Eq. 16 which too can be experimentally verified.
Also if the second term on the RHS in Eq. 19 becomes positive then $\rho_i(\textbf{r}, \gamma)$ becomes negative which can be verified by the
way one has to balance $\mu_\beta$. 
A negative number of states accommodating negatively charged electrons can behave as a positive charge cloud.
If it can attract one electron, it can also attract another electron and thus mediate an electron-electron attraction.

\bibliographystyle{References}

\end{document}